\documentclass[aps,prl,superscriptaddress,letterpaper,twocolumn,longbibliography]{revtex4-2}
\usepackage[utf8]{inputenc}
\usepackage{titlesec}
\usepackage{mwe}
\usepackage{graphicx}
\usepackage{svg} 
\usepackage{dcolumn}
\usepackage{bm}
\usepackage{bbm}
\usepackage{hyperref}

\usepackage{xcolor}
\usepackage{amsmath}
\usepackage{amssymb}
\usepackage{xcolor}
\usepackage{svg}

\usepackage{comment}
\usepackage{adjustbox}

\begin{document}

\title{Structured Interactions Drive Abrupt Transitions in the Spatial Organization of Microbial Communities}

\author{Mattia Mattei}
\affiliation{\small{Departament d'Enginyeria Inform{\`a}tica i Matem{\`a}tiques,Universitat Rovira i Virgili, 43007 Tarragona, Spain}}
\author{David Soriano Paños}
\affiliation{\small{Departament d'Enginyeria Inform{\`a}tica i Matem{\`a}tiques,Universitat Rovira i Virgili, 43007 Tarragona, Spain}}
\author{Mahantesh Halappanavar}
\affiliation
{Pacific Northwest National Laboratory, 902 Battelle Blvd, Richland, WA, 99354, USA}
\author{Alex Arenas}
\affiliation{\small{Departament d'Enginyeria Inform{\`a}tica i Matem{\`a}tiques,Universitat Rovira i Virgili, 43007 Tarragona, Spain}}
\affiliation
{Pacific Northwest National Laboratory, 902 Battelle Blvd, Richland, WA, 99354, USA}

\begin{abstract}

Bacteria possess diverse mechanisms to regulate their motility in response to environmental and physiological signals, enabling them to navigate complex habitats and adapt their behavior. Among these mechanisms, interspecies recognition enables cells to modulate their movement based on the ecological identity of neighboring species. 
Here, we introduce a model in which we assume bacterial species recognizes each other and interact via local signals that either enhance or suppress the motility of neighboring cells. Through large-scale simulations and a coarse-grained stochastic model, we demonstrate the emergence of a sharply non-linear transition driven by nucleation processes: increasing the density of motility-suppressing interactions drives the system from a fully mixed, motile phase to a state characterized by large, stationary bacterial clusters. Remarkably, in systems with a large number of interacting species, this transition can be triggered solely by altering the structure of the motility-regulation interaction matrix while maintaining species and interaction densities constant. In particular, we find that heterogeneous and modular interactions promote the transition more readily than homogeneous random ones. These results contribute to the ongoing effort to understand microbial interactions, suggesting that structured, non-random ones may be key to reproducing commonly observed spatial patterns in microbial communities.


\end{abstract}

\maketitle

\section{Introduction} Microbial communities in natural environments frequently exhibit regular spatial organization, including the formation of dense surface-attached patches, multicellular aggregates, or flocks~\cite{Cordero}. Spatial segregation among different microbial species is a pervasive phenomenon observed across diverse ecological contexts~\cite{Welch, sheth, O'Brien, Bar-Zeev}, yet the underlying factors and mechanisms responsible for generating such spatial heterogeneity are multifaceted and remain incompletely understood. Although it is well established that spatial patchiness can emerge spontaneously from growth and demographic fluctuations~\cite{Hallatschek, Golding}, comparatively less attention has been devoted to the role of active motility mechanisms in driving such spatial structuring. Indeed, a key aspect of bacterial adaptability is their capacity to regulate motility in response to a variety of external and internal signals. Mechanisms such as chemotaxis \cite{Budrene}, sensing of physical and chemical cues, and regulatory networks—including intercellular communication systems like quorum sensing (QS) \cite{Waters}—allow bacteria to modulate gene expression and adjust their movement strategies accordingly. Such regulation can also underly critical transitions between motile planktonic states and sessile, surface-attached communities, as observed during biofilm formation, a developmental process involving significant physiological and behavioral changes \cite{Davies, Lopez}. In many species, signals from QS or other pathways can lead to motility suppression, either through downregulation of flagellar genes or by promoting adhesive phenotypes that reduce mobility \cite{Daniels, Hoang, Curatolo}.

This feedback between local effects and motility reduction resembles the physics of motility-induced phase separation (MIPS) \cite{Cates}, a well-studied phenomenon in active matter systems. In MIPS, self-propelled particles segregate into dense and dilute regions purely due to the interplay between motility and local density, even in the absence of attractive interactions \cite{Redner, Fily}. While traditional MIPS models assume local steric interactions or self-regulation, biological systems often rely on nonlocal, signal-mediated communication. Recent works have begun to incorporate quorum sensing into active matter models: Ridgway et al. \cite{Ridgway} showed that QS-regulated motility can induce MIPS and oscillatory instabilities via intracellular feedback loops, while Duan et al. \cite{Duan} explored pattern formation driven by nonreciprocal quorum-sensing interactions between two species.

While most theoretical studies have focused on simplified settings involving one or two species, natural microbial communities typically consist of hundreds or even thousands of coexisting species that interact in complex and heterogeneous ways at very local scale. Understanding how the structure of these interactions shapes emergent properties such as diversity and stability remains one of the central, and still unresolved, questions in microbial ecology. This challenge is compounded by the empirical difficulty of accurately characterizing interactions that are highly variable and context-dependent \cite{suweis24}. As a result, a common theoretical strategy is to model these interactions as random matrices, with interaction strengths sampled from prescribed probability distributions \cite{May, Allesina}. However, there is also growing evidence that microbial interactions are not purely random but may exhibit structured organization. Despite their limitations, analyses of co-occurrence data have revealed that microbial interactions often display heavy-tail distributions, characterized by many taxa with few interactions and a few highly interacting species, as well as a certain degree of modularity \cite{Faust, Barberán, Peixoto}. These structural features of the interaction matrix have also been supported by experimental studies on resource competition and metabolic cross-feeding, which further confirmed the presence of modular organization in microbial interactions \cite{Goldford, Zelezniak}.

In this work, we investigate how local motility-regulation interactions influence collective motility and spatial organization in multispecies bacterial communities. We develop a minimal model in which different strains or species can locally enhance or suppress the motility of their neighbors, assuming that movement occurs on timescales much faster than demographic change or interaction with environmental resources, which are therefore neglected.
Using large-scale simulations we uncover a sharp non-linear state transition: as the density of motility-suppressing interactions increases, the system shifts from a well-mixed, motile state to a regime dominated by large, stationary clusters. In a simplified scenario involving two species, we developed a stochastic model that identifies nucleation, i.e. the formation of stable clusters driven by fluctuations, as the underlying mechanism responsible for the transition. Strikingly, when extending the model to a large number of interacting species, we find that the same transition can occur solely by modifying the topology of the motility-regulation interaction matrix, even when both species abundances and the overall two-body interactions density are kept fixed. In particular, we show that modular and heterogeneous interaction matrices significantly enhance the tendency of the system to undergo patchiness, compared to more random and homogeneous.

\section{Results}

\subsection{Microscopic Model}
We developed a microscopic simulation in which bacteria move according to run-and-tumble dynamics, consisting of ballistic motion at constant speed interrupted by random reorientations in two dimensions (see details in the Supplementary Material \cite{supplement}). In our model, the velocity of each bacterium is modulated by the local signals it receives from neighboring cells, modeled through a sigmoidal response function. Specifically, the velocity of a particle \( i \) belonging to species \( S \) is given by
\begin{equation}
v_{i\in S} = \frac{v_0}{1 + \exp\left(-\frac{1}{k} \sum_{S'=1}^{N} A_{SS'} \, \phi_{S'}(i)\right)},
\end{equation}
where \( v_0 \) is a fixed velocity, and \( k \) determines the steepness of the motility response. The interaction matrix \( A_{SS'} \) specifies how species \( S' \) influences the motility of species \( S \), with positive values promoting and negative values suppressing movement. The local density \( \phi_{S'}(i) \) denotes the concentration of species \( S' \) around particle \( i \), and is computed by counting the number of particles of species \( S' \) within a fixed interaction radius \( R \): 
\[
\phi_{S'}(i) = \frac{1}{\pi R^2} \sum_{j \in S'} M_{ij},
\]
where \( M_{ij} = 1 \) if the Euclidean distance between particles \( i \) and \( j \) is less than \( R \), and \( M_{ij} = 0 \) otherwise. We will usually consider the limit \( k \to 0 \), in which the sigmoidal response becomes sharp and effectively switches between two states: in this regime, a global negative signal from the surrounding environment almost completely suppresses bacterial motility, while in the absence of such signals the bacterium moves at least at speed \( v_0/2 \).

\subsection{Two Species Setting} 

To gain intuition about the outcomes of the motility model above, we begin by analyzing a simplified case with only two species, \( a \) and \( b \). A particularly illustrative scenario arises when bacteria of species \( a \) suppress each other’s motility but experience enhanced motility in the presence of species \( b \). In the sharp sigmoidal limit (\( k \to 0 \)), this implies that a bacterium of species \( a \) will cease its motion whenever the local density of species \( a \) exceeds that of species \( b \), and will resume movement once this condition is no longer met. This corresponds to setting the interaction matrix elements as \( A_{aa} = -1 \) and \( A_{ab} = +1 \), while species \( b \) remains unaffected by either species, assuming that always move freely at constant speed $v_0$ (\( A_{bb} = A_{ba} = +1 \)).

\begin{figure*}[t]  
    \centering
    \includegraphics[scale=0.55]{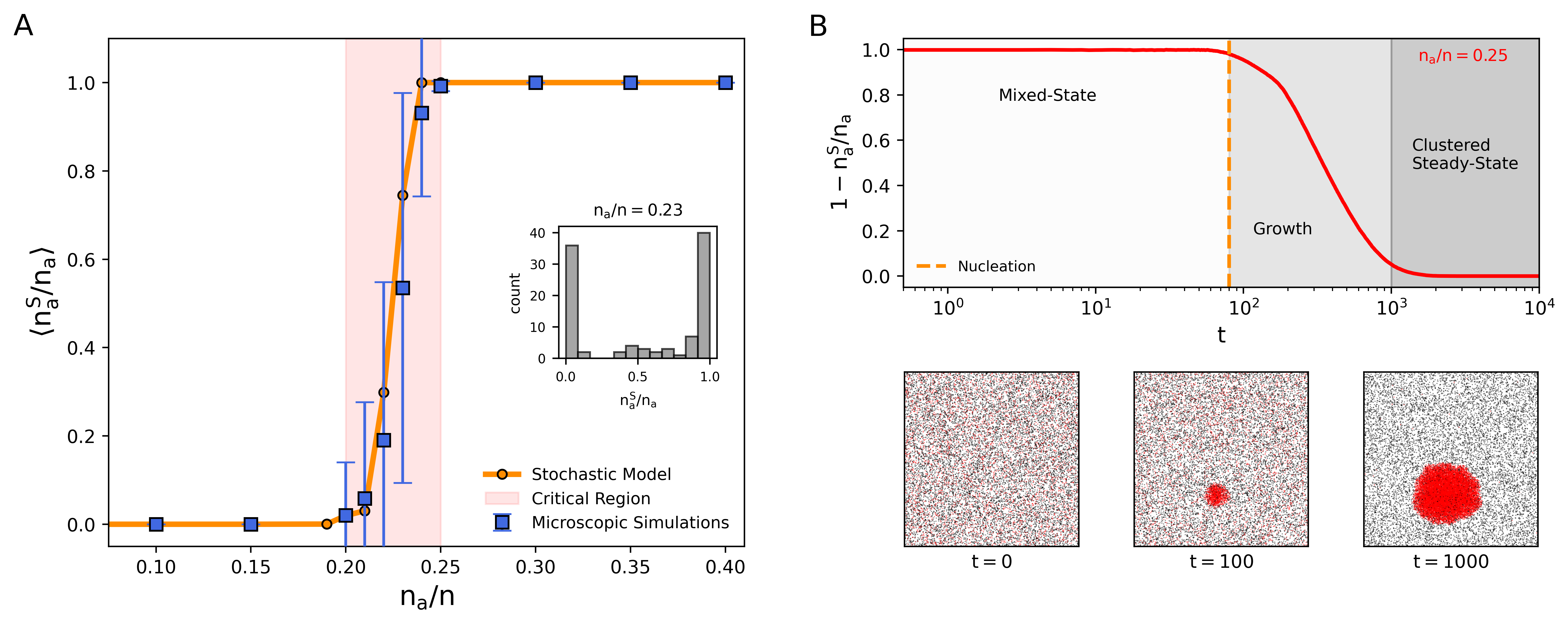}
    \makeatletter\long\def\@ifdim#1#2#3{#2}\makeatother
    \caption{(A) Mean number of non-motile bacteria at equilibrium as a function of the fraction of bacteria belonging to species ``a”. Blue squares indicate the mean values from 100 runs of the microscopic simulation, with error bars representing standard deviations. The orange line shows results obtained from the mean of 100 realizations of the theoretical stochastic model simulated using the Gillespie algorithm. The ``critical” region, where the transition may stochastically occur or not, is highlighted in slight red. The inset shows the distribution of the results for each of the 100 simulations for $n_a/n=0.23$. All the simulations here are performed with a total number of bacteria $n=16000$, an interaction radius of $R=0.025$ and a maximum velocity $v_0=0.05$. (B) Top: temporal dynamics of the fraction of motile bacteria from a representative run of the microscopic simulation at $n_a/n = 0.25$. Bottom: graphical representations of the system at three different time steps, highlighting the nucleation process. Bacteria of species ``a" and ``b" are depicted in red and black respectively.}
    \label{fig1}
\end{figure*}

In Fig.~\ref{fig1}A we present the results averaged over multiple simulations of the model, each initialized from a well-mixed configuration. The quantity plotted corresponds to the fraction of bacteria of species \( a \) that at equilibrium stop their motion. As the population of species \( a \) increases relative to species \( b \), the system exhibits an abrupt transition at a critical ratio: it shifts from a well-mixed state, where all \( a \)-type bacteria remain motile, to a phase in which large, stationary clusters of species \( a \) emerge.  Note that the intermediate points in the critical region (in red) do not reflect partial stopping within individual simulations, but rather indicate that the transition occurs only in a subset of simulations, as showed in the inset of panel A for $n_a/n=0.23$. The discontinuity of this transition is also confirmed by the presence of a pronounced hysteresis curve, which is detailed in the Supplementary Material \cite{supplement}.

Fig.~\ref{fig1}B displays the time evolution of the fraction of motile bacteria in a representative simulation performed in the critical region, showing how, after a transient phase, a nucleus spontaneously forms and attracts the accumulation of all bacteria. This phenomenology is observed across various combinations of radii and velocities (see Supplementary Material \cite{supplement}). These empirical observations of the temporal dynamics for the number of non-motile bacteria suggest that the transition is driven by a stochastic nucleation process—a mechanism widely recognized as the trigger for first-order phase transitions in condensed matter and even in biological systems as well~\cite{Kelton}. Specifically, in the vicinity of the critical region, the system exhibits a finite probability of spontaneously forming a stable, immobile cluster of species \( a \). In this context, ``stable” indicates that the cluster has reached a sufficiently high density such that the probability of bacteria within it regaining the conditions necessary to resume motility becomes very low. This localized cluster acts as a nucleation seed, triggering the irreversible arrest of motility in nearby particles and eventually propagating throughout the system. The emergence of such nuclei is a rare event in the well-mixed phase, but once formed, they grow by recruiting neighboring particles, leading to the abrupt macroscopic transition. 

To support this interpretation, we developed a coarse-grained stochastic description that explicitly accounts for the size and number of stationary clusters, along with their probabilities to form, grow, or shrink. This type of effective, mesoscopic description is commonly used in the study of nucleation phenomena~\cite{Oxtoby, Peruani}. We consider the following stochastic events: (i) \emph{Motility Loss}, $m \to s$: when a bacterium transitions from the freely moving planktonic state to a non-motile state; (ii) \emph{Motility Gain}, $s \to m$: when a stationary bacterium fulfills the conditions to resume motion; (iii) \emph{Cluster Growth}, $C_i + m \to C_{i+1}$: when a motile bacterium joins a stationary cluster of size \( i \), increasing its size to \( i+1 \); (iv) \emph{Cluster Fission}, $C_i \to C_{i-1} + m$: when a stationary cluster loses one of its bacteria, which then becomes motile again. In these processes, $C_i$ denotes a stationary cluster of size $i$ of bacteria of species $a$ (for simplicity, we omit the species index in the equations) and the notations $m$ and $s$ refer to single motile or not motile bacteria, respectively, not currently part of any cluster. The reaction rates associated with these transitions, along with the corresponding master equation, are derived in detail in the Appendix. The orange curve in Fig.~\ref{fig1}A shows the results averaged over multiple stochastic realizations of the model generated using the Gillespie algorithm~\cite{Gillespie}. These results closely reproduce the behavior observed in the microscopic simulations, thus proving that the stochastic mesoscopic description of the dynamics of clusters can sucessfully capture the emergent phenomena from the microscopic interactions.

\subsection{Large Communities}
We now generalize our formalism  to ecosystems with $N>2$ distinct species considering that the \( N \times N \) interactions matrix $A_{SS'}$ encode how species $i$ influences the motility of species $j$. This formalism permits the inclusion of specific structural properties, depending on which model is chosen to generate the matrix. To set a baseline scenario, we start by assuming that all species are equivalent and that motility-suppressing interactions are randomly distributed. This graph-construction method, known as the Erd\H{o}s–R\'enyi (ER) model~\cite{Erdos}, produces an interaction matrix lacking inherent structural correlations.
\begin{figure}
    \centering
    \includegraphics[scale = 0.55]{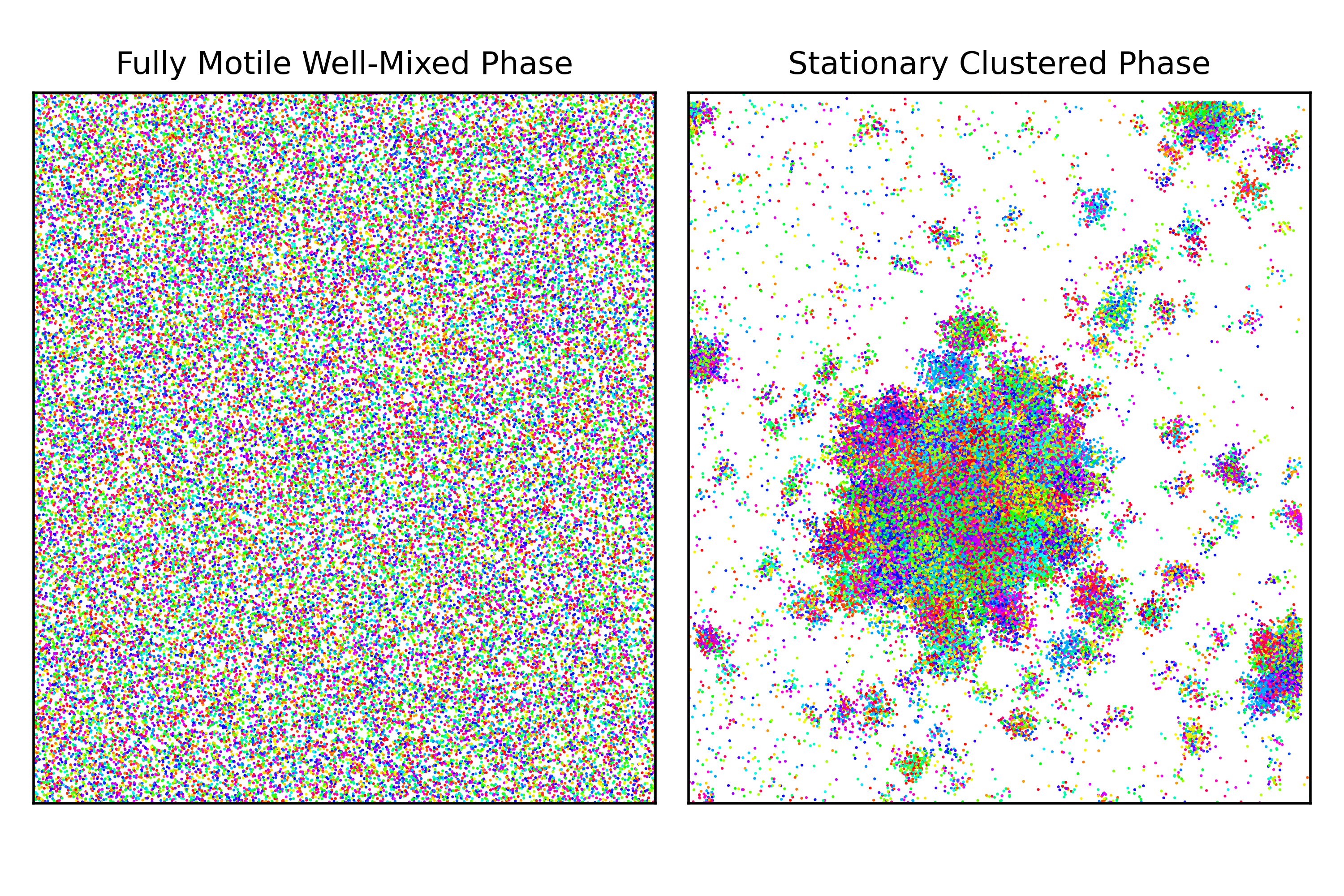}
    \caption{Steady-state spatial configurations of \( n = 100000 \) bacteria partitioned into \( N = 100 \) different species, shown for two distinct values of the number of motility suppression links. When the number of suppression links is insufficient, the system remains in a fully motile, well-mixed phase. Upon reaching a critical value, the system may undergo a transition to a stationary clustered state in which all bacteria cease motion. Simulations are performed with \( R = 0.01 \) and \( v_0 = 0.05 \) within a two-dimensional box of size \([0,1] \times [0,1]\).
}
    \label{fig2}
\end{figure}
Specifically, for a fixed suppression interactions density $L_-$, we assigned randomly $A_{SS^\prime}=-1$ when two species $S$ and $S'$ suppress each other's motility. Then, to simplify the framework, we defined its dual as representing motility enhancement, setting the remaining entries to \( +1 \). This ensures that all interactions are exclusively suppressive or enhancing. Under this configuration, simulations with \( N = 100 \) different species display phenomenology analogous to that observed in the two-species scenario, characterized by a sharp transition between a well-mixed motile phase and a stationary clustered phase, wherein bacteria of all species lose motility (see Figure~\ref{fig2}). Here, the control parameter shifts from species abundances to the density of motility suppression interactions, \( L_- / \bigl(N(N-1)/2\bigr) \). The orange curve in Fig.~\ref{fig3} shows that an increase in the proportion of suppression links drives the system toward configurations where nucleation events are more likely, consequently promoting the emergence of the transition (see the orange curve in figure \ref{fig3}).
\begin{figure}
    \centering
    \includegraphics[scale = 0.4]{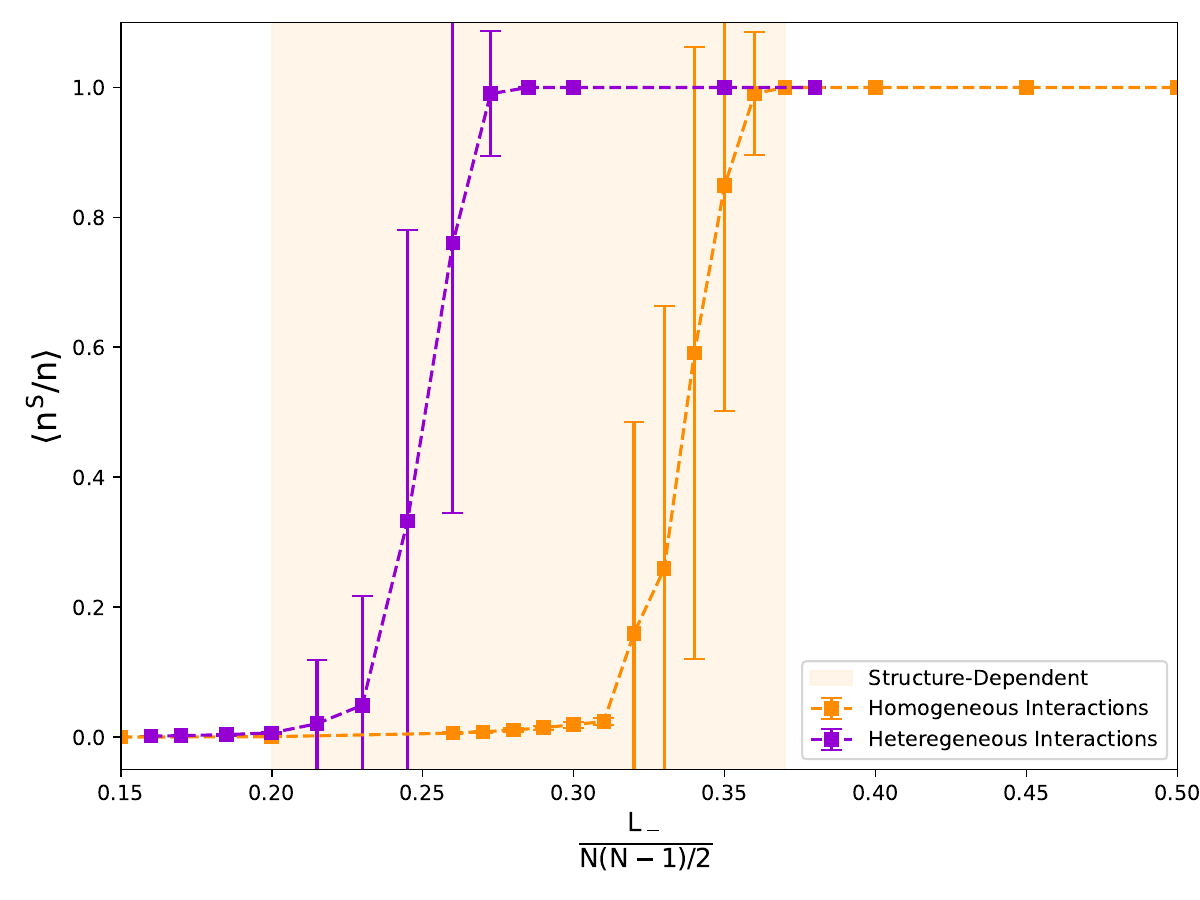}
    \caption{Mean fraction of non-motile bacteria at equilibrium $n^S/n$ as a function of the density of species-specific motility suppression interactions $2L_-/N(N-1)$. The orange and the violet curves correspond to a homogeneous interactions matrix generated by Erd\H{o}s–R\'enyi (ER) model and a heterogeneous one provided by the Barabási–Albert (BA) model. Dots indicate averages over 100 independent microscopic simulations, with error bars representing standard deviations. The yellow region highlights the bistable domain where the system’s phase depends solely on the structure of the interactions matrix. All simulations are performed within a two-dimensional box of size \([0,1] \times [0,1]\) and with \( n = 100000 \) bacteria, \( N = 100 \) species, \( R = 0.01 \), and \( v_0 = 0.05 \).
}
    \label{fig3}
\end{figure}

We then examined whether particular structures in the interactions matrix could modulate the previously observed behavior. In particular, we considered more heterogeneous matrices, where a few species engage in many interactions while the majority interact with only a few, as well as modular matrices, where species are organized into distinct groups with dense interactions within each group and sparse interactions between groups. These structured patterns in the interaction matrix are relevant because microbial systems frequently exhibit either heterogeneity or modularity in their inter-species interactions. In Figure~\ref{fig3} we present a comparison between an homogeneous matrix generated by the Erd\H{o}s–R\'enyi (ER) model and more heterogeneous ones produced using the Barabási–Albert (BA) model~\cite{Barabasi}. The BA model generates graphs with power-law degree distributions by iteratively adding nodes according to a preferential attachment mechanism, whereby new nodes are more likely to connect to existing nodes with higher degrees. Fig.~\ref{fig3} shows how, in the case of heterogeneous interactions, the critical point of the transition falls much earlier respect to the case of homogeneous ones, highlighting a region of bistability in which the system’s state—whether remaining in a motile, well-mixed phase or transitioning to a stationary clustered phase—depends solely on the structure of the motility suppression graph, rather than on the density of the interactions or species abundances.

\begin{figure}
    \centering
    \includegraphics[scale = 0.4]{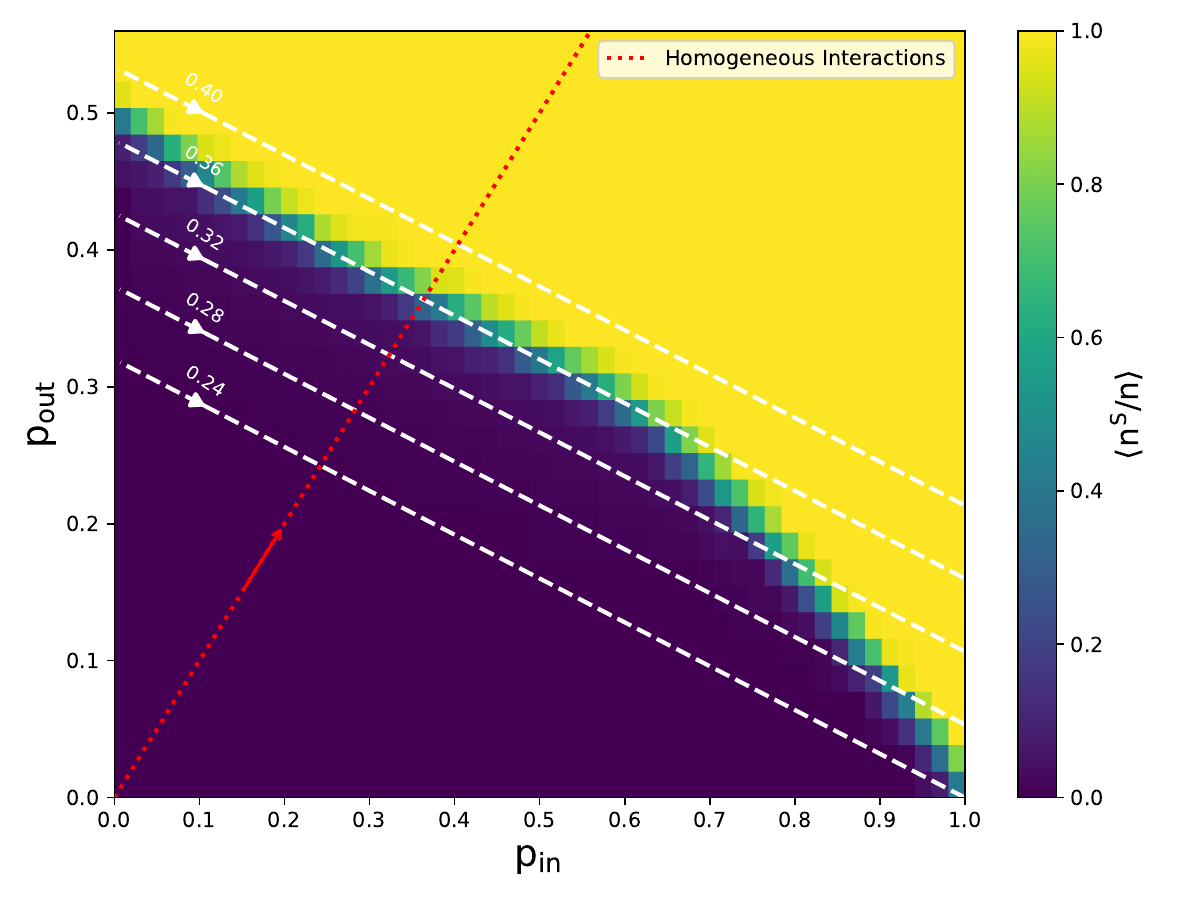}
    \caption{Heatmap showing the average number of non-motile bacteria at equilibrium for modular motility suppression matrices generated using a Stochastic Block Model. The x-axis represents \( p_{\text{in}} \), the probability of forming interactions within groups, while the y-axis corresponds to \( p_{\text{out}} \), the probability of establishing interactions between different groups. White dashed lines denote curves of constant interactions density, indicating configurations with identical numbers of interactions but different combinations of \( p_{\text{in}} \) and \( p_{\text{out}} \). The red dotted line represents the scenario of fully homogeneous interactions, where \( p_{\text{in}} = p_{\text{out}} \). A step size of 0.02 was used for both \( p_{\text{in}} \) and \( p_{\text{out}} \), and each grid cell displays the mean value computed from 100 independent microscopic simulations. All simulations were conducted within a two-dimensional domain of size \([0,1] \times [0,1]\), with \( n = 100000 \) bacteria, \( N = 100 \) species, \( R = 0.01 \), and \( v_0 = 0.05 \).
}
    \label{fig4}
\end{figure}

To investigate the impact of modular structure in the interaction matrix, we employed a simplified version of the Stochastic Block Model~\cite{Holland}. Specifically, we assume that species are partitioned into distinct communities or groups, with a probability \( p_{\text{in}} \) of establishing motility-suppressing interactions between species within the same group, and a probability \( p_{\text{out}} \) of connecting to species in other groups. Configurations where \( p_{\text{in}} \gg p_{\text{out}} \) correspond to highly modular interactions matrices, whereas the limit \( p_{\text{in}} \approx p_{\text{out}} \) recovers an homogeneous one. In Figure~\ref{fig4}, we examine the fraction of bacteria that become non-motile at equilibrium in a range of \( p_{\text{in}} \) and \( p_{\text{out}} \) values for the motility suppression matrix. Interestingly, our results indicate that interactions matrix with pronounced modularity facilitate the transition with substantially fewer suppression interactions: increasing the density of intra-group interactions while keeping the overall density of interaction constant allows transition between the motile and the stationary cluster phases. Taken together, both results show how the latter transition is given not only by the overall density of motility-suppressing interactions but also on the underlying structure of the interaction matrix.

\section{Discussion}
Several studies have underscored the importance of chemotaxis and other directed movement behaviors in reproducing specific microbial spatial patterns~\cite{Ben-Jacob, Bonner}. Moreover, recent theoretical work has hypothesized that avoidance of high-density regions could contribute to the formation of segregated bacterial clusters~\cite{Mattei}. However, microbial cells possess also the ability to differentiate between strains or species—a phenomenon known as kin discrimination~\cite{Momeni}. To our knowledge, only recent studies~\cite{Canalejo, Dinelli} have directly investigated how species-specific interactions in multi-species systems can modulate bacterial motility, although without addressing the specific structure of the underlying interaction matrix. More similar in spirit to this work is the study by Diego et al.\cite{Diego}, which specifically addresses the role of network topology in the formation of spatial patterns, although in the context of reaction-diffusion systems and Turing patterns rather than MIPS-like dynamics.

In this study, we have demonstrated that a minimal model incorporating local species-specific motility modulation can robustly reproduce patchy spatial structures commonly observed in microbial communities. Critically, our analysis identifies nucleation as a fundamental process that seeds the formation of these patterns, suggesting it as a potentially  mechanism relevant in contexts such as biofilm development~\cite{Worlitzer}. Moreover, our results provide evidence that structured interactions matrices—specifically modular and heterogeneous topologies—more effectively facilitate the emergence of optimal spatial organization compared to random, unstructured ones. This finding is especially significant given that motility-regulating interactions can serve as proxies for broader ecological interactions among species. In fact, spatial structure is widely recognized to confer significant ecological advantages. In particular, spatial segregation has been proposed as a mechanism for maintaining cooperation by fostering interactions predominantly among genetically related individuals, thereby ensuring that cooperative behaviors disproportionately benefit kin~\cite{Gore, Pande, Wu, Nadell}. In contrast, spatial mixing increases vulnerability to exploitation by non-cooperative lineages that can reap shared benefits without incurring associated costs~\cite{Nadell}. It is therefore reasonable to assume a correspondence in which mutual motility suppression reflects cooperative ecological interactions, while mutual motility enhancement represents competitive interactions. Viewed from this perspective, our results may also provide insights into the specific structures that ecological interactions might adopt, especially given the widespread patchy spatial patterns observed in microbial communities. In addition, in natural environments, nutrients are often limited, leading to widespread competitive ecological interactions. We showed that under stronger competition (modeled here as a higher number of motility-enhancing links), having a structured interaction network provides a possible mechanism for the emergence of spatial organization, which can still confer stability to the system.

The sharply non-linear transition we have observed, represents a distinctive example of motility-induced phase separation (MIPS). Unlike conventional MIPS, typically driven by changes in global density or motility parameters, our results demonstrate that similar phase transitions can emerge solely through alterations in the structure of the inter-species interactions matrix. These insights highlight the potential for matrix topology itself to act as a critical determinant of large-scale spatial organization in microbial ecosystems.

\section{Methods} 

\subsection{Coarse-Grained Stochastic Model}

We consider the simplified scenario described as the “Two-Species Setting” in the main text, consisting of two species, \( a \) and \( b \). Recall that bacteria belonging to species \( a \) can suppress each other’s motility, potentially leading to the formation of stationary clusters, whereas bacteria of species \( b \) remain motile at a constant speed \( v_0 \). 

We describe mesoscopically the system of bacteria of species \( a \) by the state vector
\[
\mathbf{n}(t) = \bigl( n_0(t), \, n_1(t), \, n_2(t), \, \ldots, \, n_{n_\text{a}}(t) \bigr),
\]
where: \( n_0 \) denotes the number of free motile bacteria, and \( n_i \) (for \( 1 \le i \le n_{n_a} \)) denotes the number of clusters composed of \( i \) non-motile bacteria. Let \( P(\mathbf{n}, t) \) denote the probability that the system occupies state \( \mathbf{n} \) at time \( t \). To define the stochastic dynamics, we identify the possible transitions between different states of the vector and their associated rates.

\emph{Motility Loss (\( 0 \to 1 \))}- Consider a motile a-type bacterium $i$. In the limit \( k \to 0 \) of the sigmoid function defining the velocity (Eq.~1 in the main text), bacteria of species \( a \) can suppress their motility when the number of nearby bacteria of species \( a \) exceeds that of the motility-enhancing bacteria of species \( b \) within a neighborhood of radius \( R \). In a well-mixed system, the numbers of species \( a \) and \( b \) bacteria inside a circle of radius \( R \) centered in the position of the bacterium $i$, denoted by \( n_a^R(i) \) and \( n_b^R(i) \), respectively, are Poisson-distributed with means \( \lambda_a = \pi R^2 n_a \) and \( \lambda_b = \pi R^2 n_b \). For the sake of clarity, from then on we will omit the notation $(i)$. 

For the bacterium $i$ to transition from a motile to a non-motile state during a time step \( \Delta t \), it must experience a change from \( n_a^R(t) \le n_b^R(t) \) to \( n_a^R(t+\Delta t) > n_b^R(t+\Delta t) \). Such a transition can occur either through the entry of additional species \( a \) bacteria into the neighborhood or through the exit of species \( b \) bacteria from the same region. If we consider sufficiently small time steps, it is sufficient to account only for events involving the entry of a single bacterium of species \( a \) into the neighborhood or the exit of a single bacterium of species \( b \). Under these conditions, the relevant transition corresponds to a change from \( n_a^R(t) = n_b^R(t) \) to \( n_a^R(t+\Delta t) = n_b^R(t+\Delta t) + 1 \). The probability that a bacterium of species \( a \) enters the neighborhood of the bacterium $i$ during the time interval \( \Delta t \) can be expressed as the product of two factors. First, the probability that another \emph{motile} bacterium of species \( a \) is located within the annular shell \([R, R + v_0 \Delta t]\) at time \( t \), and second, the probability that it moves inward across the boundary into the region of radius \( R \) around $i$. The probability of finding a motile bacterium of species \( a \) within the shell is given by the product of the density \( n_0 / A \) and the shell’s area, which is 
\[
\pi \bigl( R + v_0 \Delta t \bigr)^2 - \pi R^2 \approx 2 \pi R v_0 \Delta t + o(\Delta t^2).
\]
Given the smallness of \( v_0 \Delta t \), the probability that the bacterium moves inward across the area can be taken as \( 1/2 \), as the annular shell is restricted to the circle perimeter. Therefore, the overall probability for the entry of a single motile species \( a \) bacterium into the neighborhood during \( \Delta t \) is
\[
P_\text{entry}^a \approx n_0 \pi R v_0 \Delta t,
\]
where we recalled that $A=1$. For a bacterium of species \( b \), the probability of either entering or exiting the neighborhood during a time interval \( \Delta t \) is also given by
\[
P_\text{entry}^b = P_\text{exit}^b \approx n_b \pi R v_0 \Delta t,
\]
since these bacteria move continuously at a constant speed \( v_0 \) in random directions, so that the in- and out-fluxes are the same.

The combined probability of either a species \( a \) bacterium entering the neighborhood or a species \( b \) bacterium exiting it must be multiplied by the probability that, at time \( t \), the local populations satisfy \( n_a^R(t) = n_b^R(t) \). This probability is given by
\[
P\bigl(n_a^R = n_b^R, t\bigr)
=
\sum_{n_a^R(t)=1}^{n_a}
P\bigl(n_a^R(t)\bigr) \,
\frac{P\bigl(n_b^R(t) = n_a^R(t)\bigr)}{P\bigl(n_b^R(t)\ge n_a^R(t)\bigr)},
\]
where $P(n_a^R(t))$ and $P(n_b^R(t))$ are the Poisson distributions for the local populations of a-type and b-type bacteria. The division by \( P\bigl(n_b^R(t) \ge n_a^R(t)\bigr) \) arises because \( n_b^R(t) \) must be at least equal to \( n_a^R(t) \); otherwise, bacterium $i$ would already have transitioned to a non-motile state. Substituting the Poisson distributions, we obtain
\[
P\bigl(n_a^R = n_b^R, t\bigr)
=
\sum_{n_a^R(t)=1}^{n_a} \sum_{n_b^R\ge n_a^R}
\frac{e^{-\lambda_a} \, \lambda_a^{n_a^R} \lambda_b^{n_a^R-n_b^R}\, n_b^R!}
{\bigl(n_a^R!\bigr)^2},
\]
where, for clarity, the time argument has been omitted in the right term of the equation.
Finally the total rate, $W_{0\to1}$, for unit of time at which free bacteria of species $a$ loose motility is just given by

\begin{align}
    W_{0 \to 1} 
    &= \frac{n_0}{\Delta t} \,
    \Bigl(
        P\bigl(n_a^R = n_b^R\bigr) \,
        \bigl( P_\text{entry}^a + P_\text{exit}^b \bigr)
    \Bigr) \notag \\
    &= n_0 \pi R v_0 \,
    \Bigl(
        P\bigl(n_a^R = n_b^R\bigr) \,
        \bigl( n_0 + n_b \bigr)
    \Bigr).
\end{align}

\emph{Motility Gain (\(1 \to 0\))}- The derivation of the motility gain rate follows a similar procedure as in the previous section. Here, the relevant transition during a sufficiently small time step is the change from \( n_a^R(t) = n_b^R(t) + 1 \) to \( n_a^R(t + \Delta t) = n_b^R(t + \Delta t) \). This transition can occur either through the entry of a new bacterium of species \( b \) into the circle of radius \( R \) surrounding the considered non-motile bacterium \( i \), or through the exit of a free bacterium of species \( a \) from that region. Therefore, we consider the probability
\begin{align}
P\bigl(n_a^R = n_b^R + 1, t\bigr)
&= \sum_{n_a^R(t)=1}^{n_a}
P\bigl(n_a^R(t)\bigr) \notag \\
&\quad \times 
\frac{P\bigl(n_b^R(t) = n_a^R(t) - 1\bigr)}
     {P\bigl(n_b^R(t) < n_a^R(t)\bigr)}. \notag
\end{align}
where the denominator serves as a normalization factor because, at time \( t \), it must hold that \( n_b^R(t) < n_a^R(t) \) for bacterium \( i \) to remain in a non-motile state. The total transition rate \( W_{1 \to 0} \), representing the probability per unit time that non-motile bacteria of species \( a \) regain motility, is then given by

\begin{align}
    W_{1 \to 0} 
    &= \frac{n_1}{\Delta t} \,
    \Bigl(
        P\bigl(n_a^R = n_b^R + 1\bigr) \,
        \bigl( P_\text{entry}^b + P_\text{exit}^a \bigr)
    \Bigr) \notag \\
    &= n_1 \pi R v_0 \,
    \Bigl(
        P\bigl(n_a^R = n_b^R + 1\bigr) \,
        \bigl( n_0 + n_b \bigr)
    \Bigr).
\end{align}

\emph{Cluster Growth (\(C_i \to C_{i+1}\))}- We assume that each bacterium occupies a circular area with diameter \( d \), representing its physical size and that clusters are perfect circles with radius \( R_C \). Thus, the total area occupied by \( i \) bacteria is given by $A(i) = i \pi d^2/4$. If \( i \) non-motile bacteria of species \( a \) are arranged in an optimal hexagonal close packing, the cluster can be approximated by a single circular region with effective area $A_{C_i} = A(i)/\eta$, where \(\eta = \pi / (2\sqrt{3})\) denotes the packing efficiency of hexagonal close packing. The corresponding cluster radius is therefore $R_{C_i} = d \sqrt{i \sqrt{3}/{2 \pi}}$. For a cluster to grow, a motile bacterium of species \( a \) must collide with the interaction zone surrounding the cluster. This interaction area has a radius \( R + R_{C_i} \), yielding a total cross-sectional area of \(\pi (R + R_{C_i})^2\). As before, we consider small time steps \(\Delta t\) and assume only one collision event per interval. The relevant shell region is defined by \([R + R_{C_i},\, R + R_{C_i} + v_0 \Delta t]\), whose area is approximately $2 \pi v_0 \Delta t \bigl( R + R_{C_i} \bigr) + o(\Delta t^2)$. Assuming that the probability for a bacterium in this shell to move inward is again \( 1/2 \), the probability that a motile bacterium of species \( a \) enters the interaction region during \(\Delta t\) is given by $P_\text{entry}^a =
n_0 \pi v_0 \bigl( R + R_{C_i} \bigr) \Delta t$. The corresponding total cluster growth rate per unit time, \( W_{C_i \to C_{i+1}} \), for clusters of size \( i \) is therefore

\begin{align}
W_{C_i \to C_{i+1}} 
&= n_i \, n_0 \, \pi v_0 \bigl( R + R_{C_i} \bigr) \notag \\
&= n_i \, n_0 \, \pi v_0 
\left(
    R + d \sqrt{\frac{i \sqrt{3}}{2 \pi}}
\right).
\end{align}

\emph{Cluster Fission ($C_i\to C_{i-1}$)}- Clusters can subsequently lose bacteria due to new collisions with \( b \)-type bacteria. Consider a bacterium situated at the perimeter of a cluster with radius \( R_{C_i} \). If \( R_{C_i} \le R/2 \), this bacterium interacts with the entire cluster, otherwise, it interacts only with a fraction \(\alpha\) of it. To compute \(\alpha\), we calculate the area of intersection between the circular interaction zone of radius \( R \) centered on the edge bacterium and the cluster area \( A_{C_i} \). This intersection area is given by \cite{URL}

\begin{align}
A_{\cap} 
&= R_{C_i}^2 \cos^{-1} 
\left(
    1 - \frac{R^2}{2 R_{C_i}}
\right) 
+ R^2 \cos^{-1}
\left(
    \frac{R}{2 R_{C_i}}
\right) \notag \\
&\quad - \frac{1}{2} R \sqrt{4 R_{C_i}^2 - R^2}. \notag
\end{align}

so that we approximate $\alpha = A_{\cap}/A_{C_i}$, when $R_{C_i}>R/2$, otherwise $\alpha = 1$. Thus, a bacterium at the cluster boundary experiences a local number of suppression interactions $n_a^R = \alpha i + n_a^{R,o}$, where \( n_a^{R,o} \) represents interactions with other motile \( a \)-type bacteria located outside the cluster. Consider the case where, at time \( t \), the condition $n_B^R(t) = n_a^R(t) - 1 = \alpha i + n_a^{R,o} - 1$
holds. If a \( b \)-type bacterium collides with the area of radius \( R_{C_i} + R \), or if a free \( a \)-type bacterium exits the shell defined by \([R_{C_i}, R_{C_i} + R]\), the bacterium at the boundary can regain motility. Using a reasoning analogous to previous derivations, the total rate at which clusters of size \( i \) lose bacteria is given by
\begin{align}
W_{i \to i-1} &= n_i \pi v_0 (R_{C_i} + R) \sum_{n_a^{R,o}} P(n_a^{R,o}) \notag \\
&\quad \times \frac{P(n_B^R = n_a^{R,o} + \alpha i - 1)}{P(n_B^R < n_a^{R,o} + \alpha i)} (n_0 + n_B).
\end{align}

where \( n_a^{R,o} \) is assumed to be Poisson-distributed, exactly as \( n_a^R \).

\emph{Master Equation}—Thus the time evolution of \( P(\mathbf{n}, t) \) can be described by
\begin{align}
\frac{d}{dt} P(\mathbf{n}, t) 
&= \sum_r \big[
W_r(\mathbf{n} - \Delta \mathbf{n}_r)\,
P(\mathbf{n} - \Delta \mathbf{n}_r, t) \notag \\
&\quad - W_r(\mathbf{n})\, P(\mathbf{n}, t)
\big].
\end{align}

where \( r \in \{0 \to 1,\; 1 \to 0,\; C_i \to C_{i\pm1}\} \), with corresponding updates to \( \mathbf{n} \). This stochastic dynamics is solved via the Gillespie algorithm, yielding trajectories that match the microscopic simulations (Fig.~1 and SM).

\section{Acknowledgments}
Work supported by Spanish Ministerio de Ciencia e Innovaci\'on (PID2021-128005NB-C21), Generalitat de Catalunya (2021SGR-00633), Universitat Rovira i Virgili (2023PFR-URV-00633) and the European Union Horizon Europe Programme under the CREXDATA project (grant agreement no.\ 101092749). AA acknowledges ICREA Academia, and the Joint Appointment Program at Pacific Northwest National Laboratory (PNNL). PNNL is a multi-program national laboratory operated for the U.S.\ Department of Energy (DOE) by Battelle Memorial Institute under Contract No.\ DE-AC05-76RL01830. The project has received also funding from the European Union’s Horizon 2020 research and innovation program under the Marie Skłodowska-Curie grant agreement No. 945413 and from the Universitat Rovira i Virgili (URV). Disclaimer: This work reflects only the author’s view and the Agency is not responsible for any use that may be made of the information it contains.

\section{Author Contributions}
MM, DSP and AA designed the study. MM and MH wrote code. MM wrote equations and performed the analysis. All the authors analyzed results, discussed results, and wrote the paper.

\section{Competing Interests}
The authors declare no competing financial interests.

\bibliographystyle{apsrev4-2}
\bibliography{bibliography.bib}

\clearpage

\onecolumngrid      

\section*{Supplementary Material}

\subsection*{S1) Spatial Organization and Cluster Size Distributions}

In this section, we further investigate the final spatial organization of bacterial communities arising from different structures of interaction matrices, namely homogeneous, heterogeneous, and modular configurations, analyzed in the main text. For each scenario, we explore various values of suppression link connectivity, ranging from the critical region to the super-critical regime.

To characterize these spatial organizations, we examine the statistical distribution of local bacterial densities at the stationary state. This is achieved by partitioning the spatial domain into a fine grid and counting the number of bacteria in each cell. A natural choice for the cell size is the interaction radius, which in all cases is set to \( R = 0.01 \).

\begin{figure}[h]
    \centering
    \includegraphics[scale = 0.35]{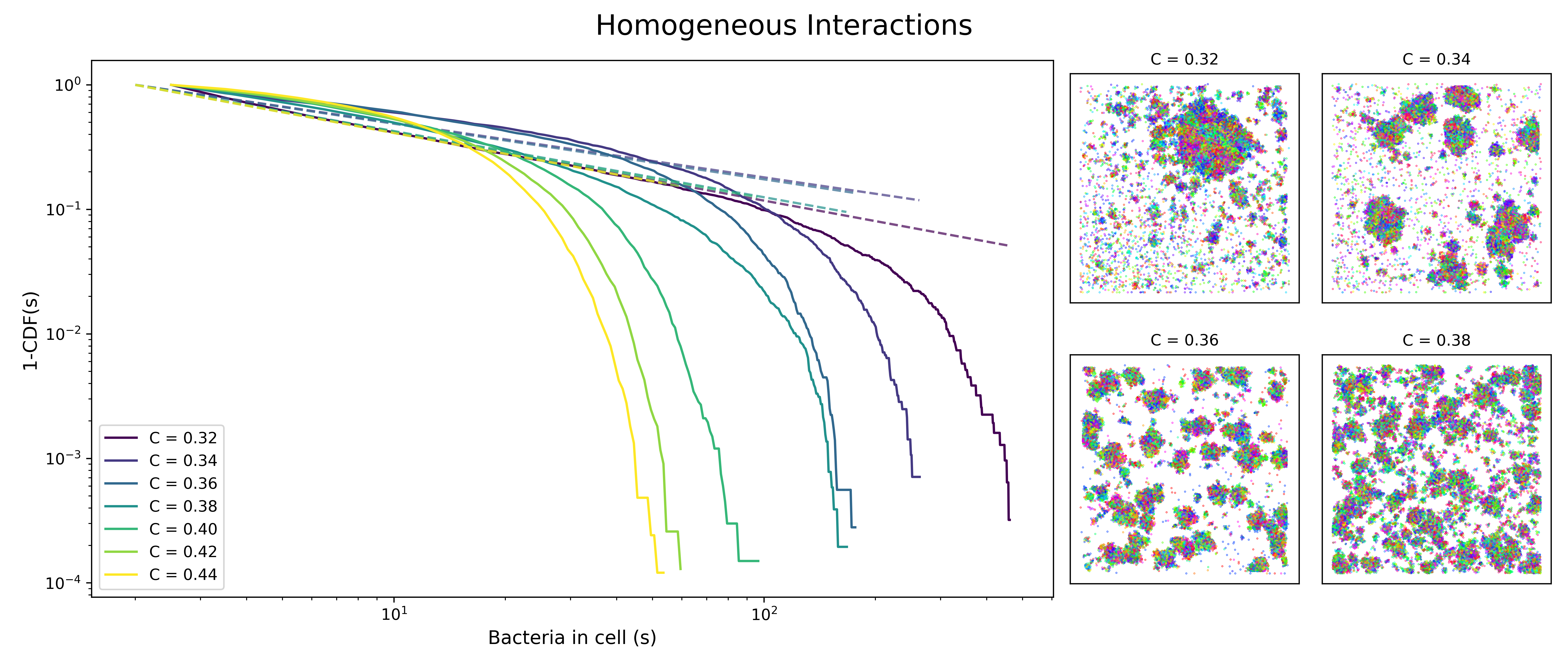}
    \caption{On the left, the complementary cumulative distribution function (CCDF) for the number of bacteria in a cell of size \( R \times R \) (\( R = 0.01 \)). The different solid curves correspond to various densities of suppression interactions, spanning from the critical to the super-critical regime (see Fig.~3 in the main text). Dashed lines indicate the best fits using a discrete power-law probability distribution. On the right, stationary spatial patterns are displayed for \( C = 0.32, 0.34, 0.36, 0.38 \). Each color represents a distinct species and the homogeneous interactions matrix is generated according to the Erd\H{o}s-Renyi model. All simulations are performed in a \([0,1] \times [0,1]\) domain with \( n = 100000 \), \( N = 100 \) species, interaction radius \( R = 0.01 \), and velocity \( v_0 = 0.05 \).}
    \label{figS1}
\end{figure}

In Fig.~\ref{figS1}, we first consider the case of homogeneous random interactions generated by the Erd\H{o}s-Renyi model \cite{Erdos}. It is evident that increasing the density of suppression links leads to the formation of more numerous and smaller clusters compared to the early stages of the critical region (\( C = 0.32, C = 0.34 \)), where nucleation events are less likely. From these cell occupancy counts, we compute the empirical complementary cumulative distribution function (CCDF), focusing on cells containing at least one bacterium. We plot these CCDF curves for different connectivity values \( C \). To further analyze cluster-size distributions, we fit a discrete power-law model to the occupancy data. This fitting procedure helps assess whether cluster sizes exhibit scale-invariant behavior under different connectivity regimes. However, the power-law fit tends to hold only up to a certain cluster size, beyond which the distributions drop off more sharply as \( C \) increases.

\begin{figure}[h]
    \centering
    \includegraphics[scale = 0.35]{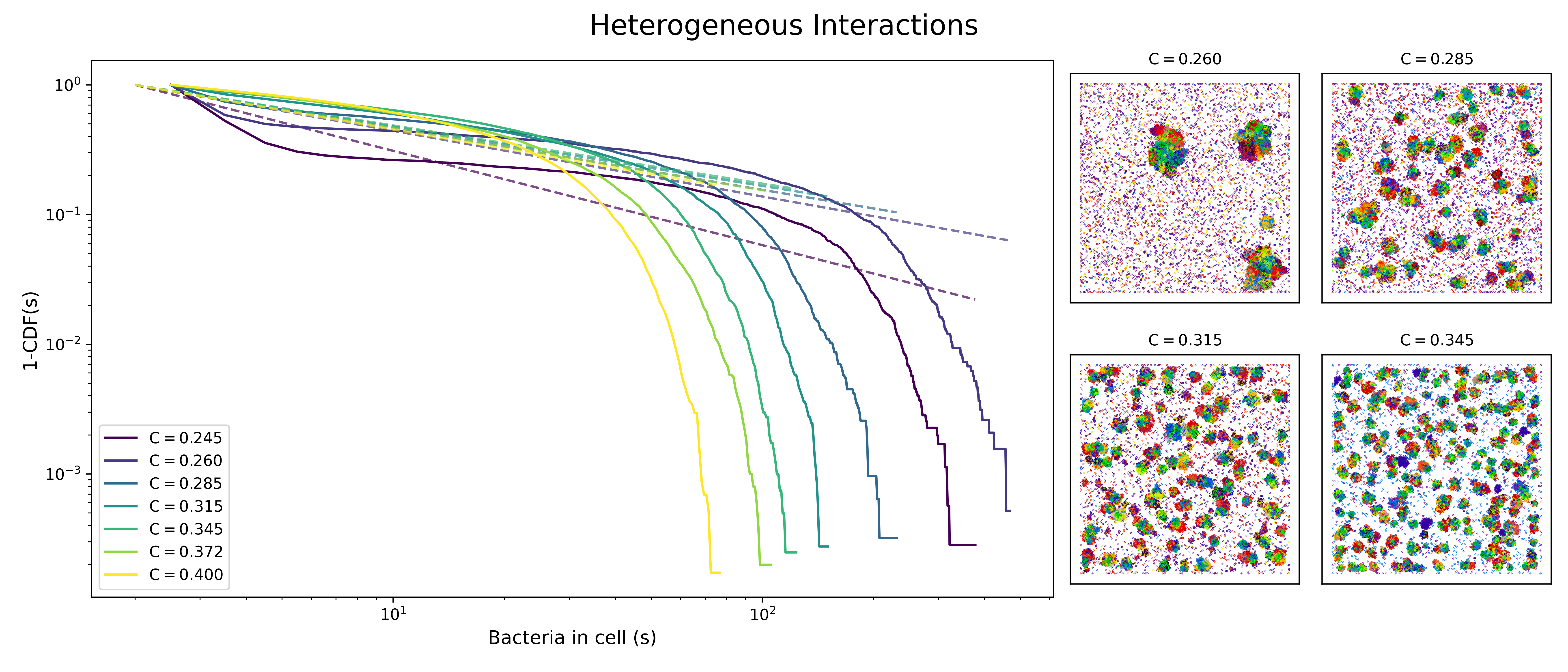}
    \caption{On the left, the complementary cumulative distribution function (CCDF) for the number of bacteria in a cell of size \( R \times R \) (\( R = 0.01 \)). The different solid curves correspond to various densities of suppression interactions, spanning from the critical to the super-critical regime (see Fig.~3 in the main text). Dashed lines indicate the best fits using a discrete power-law probability distribution. On the right, stationary spatial patterns are displayed for \( C = 0.26, 0.285, 0.315, 0.345 \). Each color represents a distinct species and the heterogeneous interactions matrix is generated according to the Barabási-Albert model. All simulations are performed in a \([0,1] \times [0,1]\) domain with \( n = 100000 \), \( N = 100 \) species, interaction radius \( R = 0.01 \), and velocity \( v_0 = 0.05 \).}
    \label{figS2}
\end{figure}

In Fig.~\ref{figS2}, we analyze the case of heterogeneous interactions generated via the Barabási–Albert model \cite{Barabasi}. Although we again observe more numerous and smaller clusters as \( C \) increases, the clusters in this scenario appear more compact and denser than in the homogeneous case. Moreover, the CCDF curves exhibit a slightly non-monotonic behavior, particularly at small cluster sizes.

\begin{figure}[h]
    \centering
    \includegraphics[scale = 0.35]{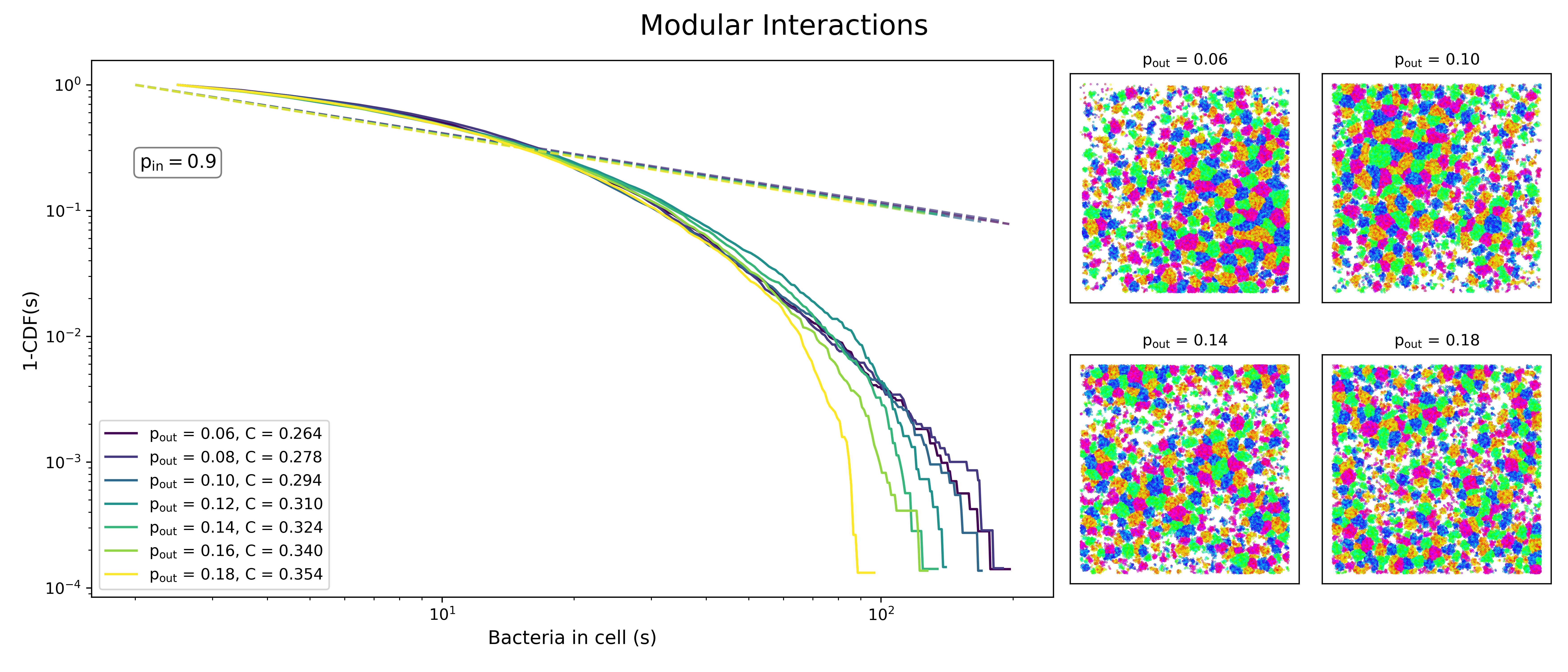}
    \caption{On the left, the complementary cumulative distribution function (CCDF) for the number of bacteria in a cell of size \( R \times R \) (\( R = 0.01 \)). The different solid curves correspond to various densities of suppression interactions obtaining using the Stochastic Block Model with fixed $p_\text{in}=0.9$ and varying $p_\text{out}$ from 0.06 to 0.18, spanning from the critical to the super-critical regime (see Fig.~4 in the main text). Dashed lines indicate the best fits using a discrete power-law probability distribution. On the right, stationary spatial patterns are displayed for \( C = 0.264, 0.294, 0.324, 0.354 \). Each color represents a distinct species. All simulations are performed in a \([0,1] \times [0,1]\) domain with \( n = 100000 \), \( N = 100 \) species, interaction radius \( R = 0.01 \), and velocity \( v_0 = 0.05 \).}
    \label{figS3}
\end{figure}

The most distinct behavior arises in the case of modular interactions, as shown in Fig.~\ref{figS3}. Here, we generate the interaction matrix using a Stochastic Block Model \cite{Holland} with \( p_\text{in} = 0.9 \) and vary \( p_\text{out} \) to increase the overall density of suppression links and transition from the critical to the super-critical regime (see Fig~4 in the main text). We chose a high value for \( p_\text{in} \) to specifically explore strongly modular structures. In this case, the spatial organization becomes more patchy and fragmented, rather than forming spatially segregated clusters as seen in the previous scenarios. The CCDF distributions in the modular case also differ markedly from those observed earlier, showing even greater deviations from the fitted power-law behavior.

These analyses highlight how strongly the structure of the interaction matrix influences the stationary spatial organization of microbial communities. This connection suggests promising directions for developing inference methods that could utilize spatial imaging data to deduce interspecies interaction networks.

\subsection*{S2) Hysteresis Curve and Clusters Stability}

Hysteresis curves are often indicative of abrupt, first-order phase transitions. To verify that the transition we observe is indeed discontinuous, we explore the stability of the stationary clustered state under perturbations that push the system in the opposite direction along the control parameter axis. In the two-species scenario, this entails decreasing the fraction of a-type bacteria relative to the total population—for example, by increasing the number of b-type bacteria, which might destabilize existing clusters of a-type cells.

We focus on several initial conditions corresponding to stationary spatial distributions at fractions \( n_a/n = 0.25, 0.30, 0.35 \), and \( 0.40 \), spanning the critical and super-critical regimes. To probe the system’s resilience, we apply a ``shock'' perturbation by introducing additional b-type bacteria and then resume the microscopic simulations to observe how the spatial organization evolves in response.

\begin{figure}
    \centering
    \includegraphics[scale = 0.5]{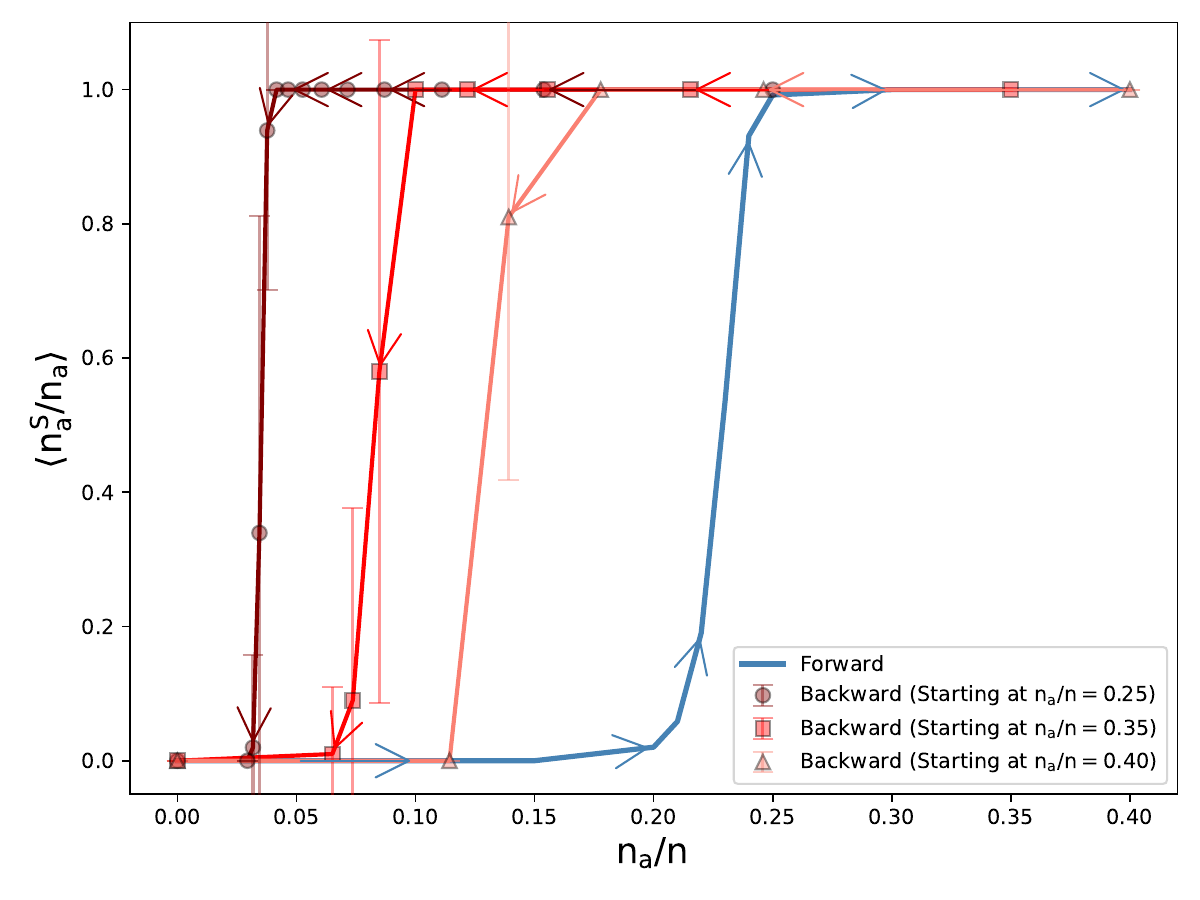}
    \caption{Mean fraction of non-motile a-type bacteria at stationarity as a function of the initial fraction \(n_a/n\), averaged over 100 microscopic simulations. The blue curve corresponds to the ``forward'' process (also shown in Fig.~1 of the main text), where the system evolves by progressively increasing the fraction of a-type bacteria. The other three curves correspond to ``backward'' perturbations, starting from stationary configurations obtained at \(n_a/n = 0.25\) (end of critical region), \(n_a/n = 0.35\), and \(n_a/n = 0.40\) (super-critical regime). For each case, b-type bacteria are introduced in increasing amounts in steps of \(10000\), with the system reset to the original unperturbed configuration before each addition.}
    \label{figS4}
\end{figure}

Our results in Fig.~\ref{figS4} reveal a pronounced hysteresis effect: significantly more b-type bacteria are required to drive the system back into the fully motile, well-mixed state than would be needed to trigger clustering in the forward direction. Interestingly, the critical fraction for this reverse transition is lower as we move from the critical into the super-critical regime. This occurs because, in the critical region, clusters are larger and denser, allowing bacteria to reorganize more readily into stable, even denser aggregates following perturbation. In contrast, in the super-critical region, clusters are smaller and the spatial organization resembles a more homogeneous, well-mixed state, making it easier for perturbations to break apart the smaller aggregates and return the system to a motile phase.

Besides confirming the sharpness of the observed transitions, these intriguing results also highlight how an initially less favorable condition, characterized by a lower number of a-type bacteria, can paradoxically lead to a more stable spatial organization compared to scenarios with a higher initial fraction of a-type cells. This counterintuitive behavior has potential implications for understanding biofilm formation and, in particular, the resilience of biofilms to external perturbations or interventions.

\subsection*{S3) Sensitivity Analysis}

To investigate whether the observed phenomenology depends on the specific parameters chosen or holds more generally, we performed a sensitivity analysis by varying the physical parameters of the model, in particular the interaction radius $R$ and the maximum velocity $v_0$. For simplicity, we carried out this analysis in the two-species scenario, as our goal was to verify whether the nucleation-driven abrupt transition persists and whether the assumptions made in our mesoscopic model remain valid across different combinations of $R$ and $v_0$.

\begin{figure}
    \centering
    \includegraphics[scale=0.5]{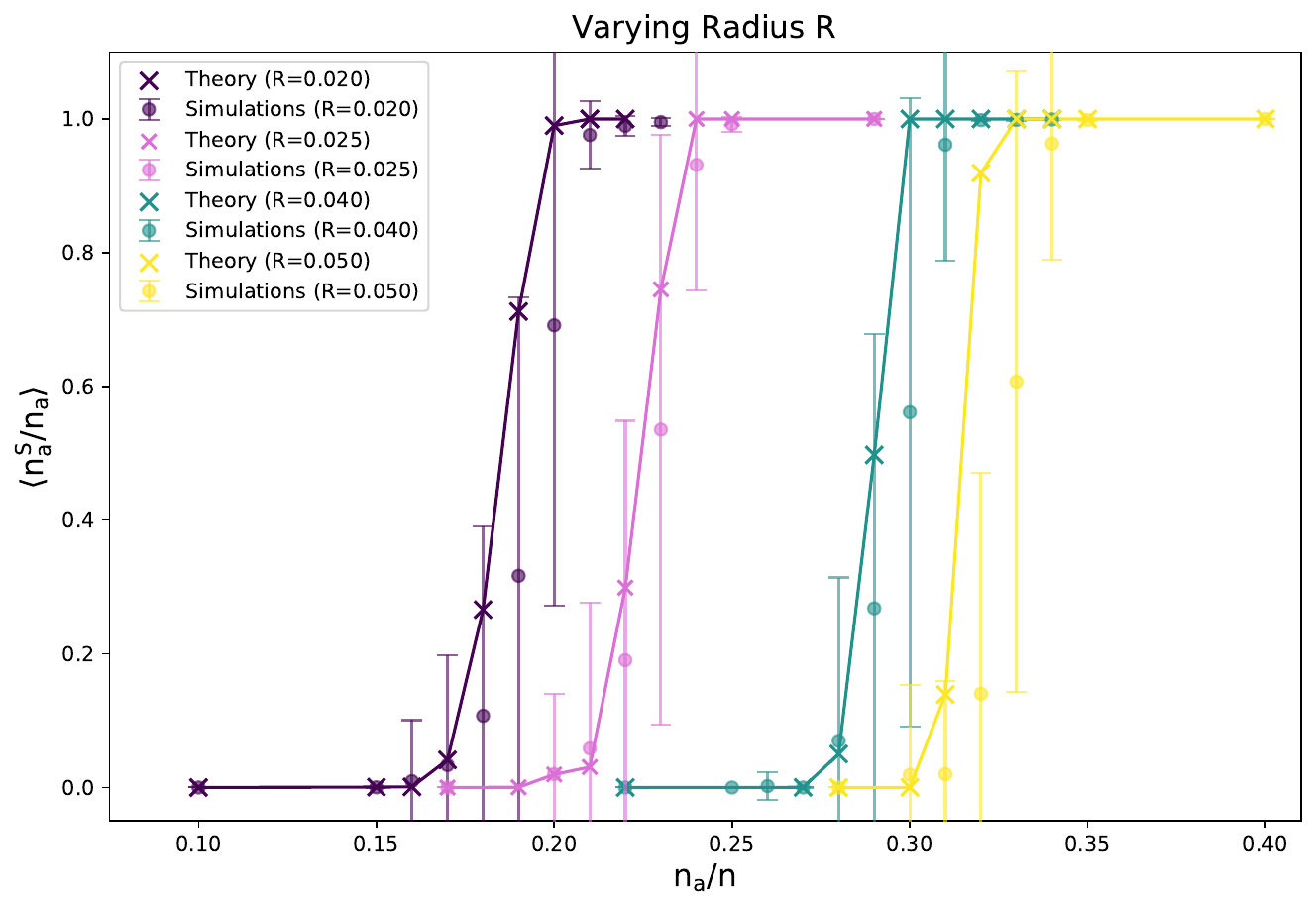}
    \caption{Mean fraction of non-motile bacteria at the stationary state as a function of the fraction of \textit{a}-type bacteria in the system. The averages are computed over 100 realizations of the microscopic simulations (depicted as small circles) and over 100 runs of the stochastic model (shown as small crosses). Both theory and simulations are evaluated for interaction radii $R = 0.02$, $0.025$, $0.040$, and $0.050$. Error bars indicate standard deviations. All simulations were performed with $n = 16{,}000$ particles, $N = 2$ species, and a maximum velocity $v_0 = 0.05$, within a spatial domain $[0,1] \times [0,1]$.}
    \label{figS5}
\end{figure}

In Fig.~\ref{figS5}, we show the transition curves for four different values of the interaction radius: $R = 0.02$, $0.025$, $0.04$, and $0.05$. We avoided exploring too small or too large radii because, in the former case, the average number of interacting bacteria becomes too low, while in the latter, the presence of too many interacting partners may not be well captured by our Poisson distribution assumption, which neglects spatial correlations.

As before, we plot the mean fraction of non-motile bacteria of species $a$ in the stationary state, averaged over 100 independent realizations, both for the microscopic simulations and for the stochastic model solved via the Gillespie algorithm. In all cases, we observe an abrupt transition. Interestingly, the critical point of the transition shows a clear dependence on the interaction radius: as $R$ increases, the transition occurs at higher values of the fraction of $a$-type bacteria in the system. This behavior arises because shorter interaction radii lead to fewer neighbors on average, which enhances fluctuations. Consequently, the likelihood of generating the local fluctuations necessary to trigger the nucleation and growth of stationary clusters increases. This highlights the biological relevance for bacteria of maintaining short-range interactions. Overall, even under these varying conditions, our stochastic model successfully reproduces the simulation results, closely capturing the two-phase behavior observed in the system.

\begin{figure}
    \centering
    \includegraphics[scale=0.4]{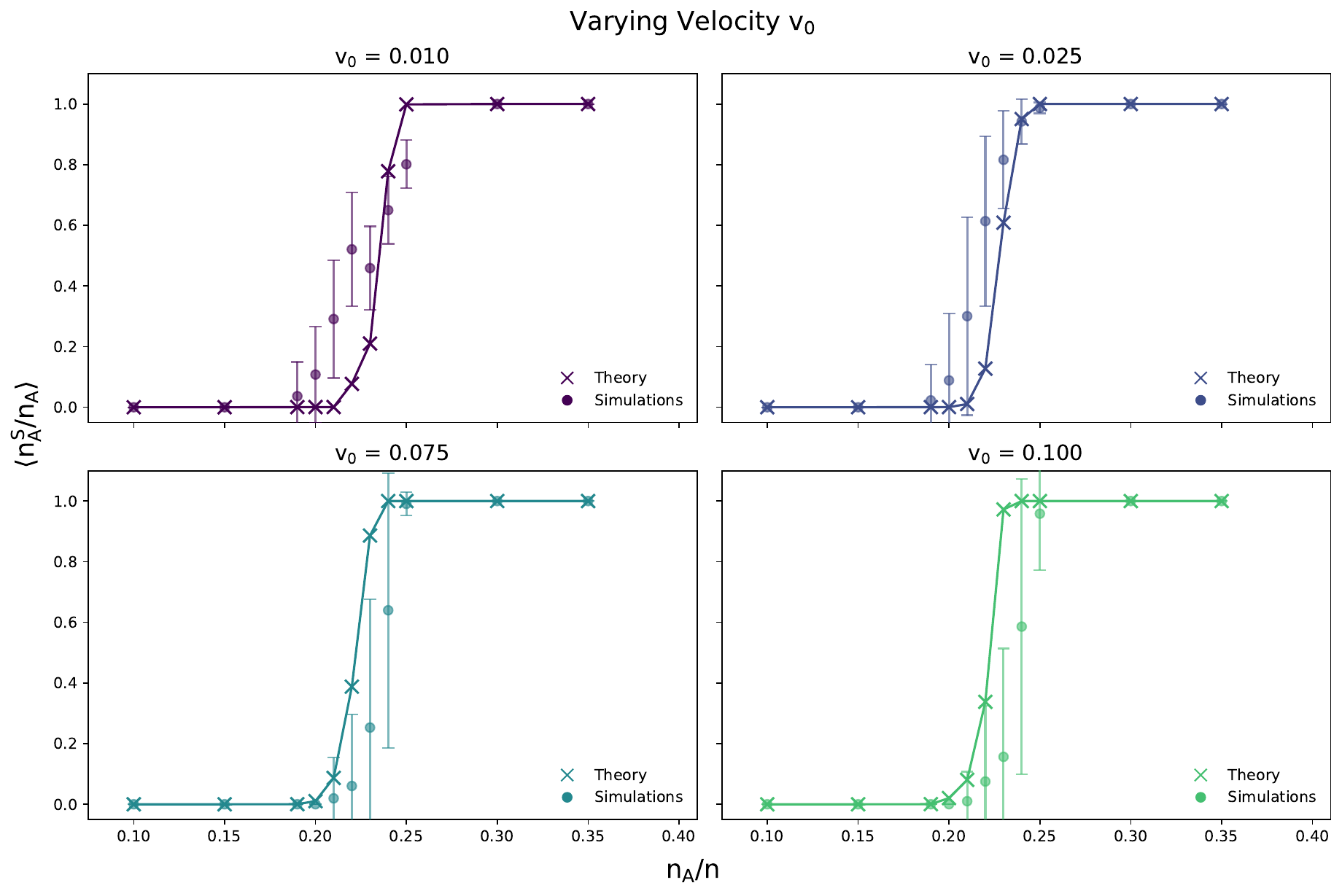}
    \caption{Mean fraction of non-motile bacteria at the stationary state as a function of the fraction of \textit{a}-type bacteria in the system. The averages are computed over 100 realizations of the microscopic simulations (depicted as small circles) and over 100 runs of the stochastic model (shown as small crosses). Both theory and simulations are evaluated for maximum velocities $v_0 = 0.01$, $0.025$, $0.075$, and $0.1$. Error bars indicate standard deviations. All simulations were performed with $n = 16{,}000$ particles, $N = 2$ species, and an interaction radius $R= 0.025$, within a spatial domain $[0,1] \times [0,1]$.}
    \label{figS6}
\end{figure}

In Fig.~\ref{figS6}, we explore the effect of varying the maximum velocity $v_0$ while keeping the interaction radius fixed. Again, we avoid considering too small or large velocities and we consider $v_0=0.01, 0.025, 0.075$ and $0.1$. As in previous analyses, we observe abrupt transitions between the fully motile, well-mixed phase and the phase characterized by stationary clusters, along with a good agreement between the stochastic model and the microscopic simulations. Interestingly, within this range of velocities, the critical point of the transition and the overall phenomenology remain largely unaffected by changes in $v_0$. Empirical observations indicate that the primary influence of velocity is on the timescale: lower velocities tend to slow down, while higher velocities accelerate, the collective arrest of the bacterial population following nucleation.

\subsection*{S4) Parameters Choice and Simulation Details}

For simulations involving $N=100$ species, parameter values were selected to reflect biologically plausible conditions. Microbial communities are typically highly dense and populated; therefore, we considered a relatively large population size of $n = 100000$. Although this is still lower than real microbial ecosystems, it offers a practical compromise between biological realism and computational feasibility, given the thousands of simulations performed in this study.

Bacterial interactions generally occur over very short distances, often spanning only a few micrometers \cite{DalCo}. Assuming our simulated spatial domain of $[0,1] \times [0,1]$ represents an area of $1\,\text{mm}^2$, we chose an interaction radius of $R = 0.01$, which corresponds to $10\,\mu\text{m}$, approximately ten bacterial body lengths. Regarding motility, bacteria exhibit a wide range of swimming speeds from $2$ to $200\,\mu\text{m/s}$ \cite{McGrawHill1960}. As a biologically reasonable intermediate and a commonly cited reference value \cite{Tortora1995, WorldBook1973}, we set the velocity to $v_0 = 0.05$, equivalent to $50\,\mu\text{m/s}$. The coefficient \( k \) appearing in the formulation of the velocity \( v_i \) (Eq.~1 in the main text) is set to \( k = 0.001 \) throughout all analyses. This value is chosen to be sufficiently small to ensure that the sigmoid function governing the velocity response remains sharp.

In the two-species case, we reduced the total number of bacteria to $n = 16000$. This adjustment ensures that the theoretical model, which depends on rates proportional to the number of $a$-type bacteria, remains computationally manageable. To maintain the same average number of interacting neighbors as in the $N=100$ case, we proportionally increased the interaction radius to $R = 0.025$.

In the model, there is a single free parameter, $d$, representing the “physical” size associated with an individual bacterium. Throughout our analyses, we set $d = 0.001$, which is sufficiently small compared to the interaction radius in the two-species scenario (ranging from 0.02 to 0.05). We verified that the results remain consistent for different values of $d$, provided that $d$ stays much smaller than $R$.

All simulations, both the microscopic and the Gillespie-based ones, incorporate a stopping criterion: the coefficient of variation of the number of non-motile bacteria must fall below 0.001 over the last 500 time steps. If this condition is not satisfied, the simulation is terminated after a total of 5000 time steps.

The microscopic simulation scripts are implemented in C++. To efficiently compute the number of interacting bacteria, the algorithm constructs a data structure known as a Random Geometric Graph (RGG)~\cite{Penrose2003}, which is updated in parallel at each time step. The parallelization leverages spatial discretization, dividing the domain into squares of side length \( R \). To identify potential neighbors, each bacterium only checks distances to others located within its own cell and the surrounding eight neighboring cells.

To avoid introducing correlations in bacterial trajectories, tumbling times \( t \)—the times at which bacteria change their movement direction—are sampled from an exponential distribution given by \(p(t) = e^{-t/\tau_0}\) with \(\tau_0 = 1\). At each simulation time step, the number of neighbors for each bacterium is first computed, which determines its instantaneous velocity \( v_i \). A new fully random movement direction is then selected. If the next tumbling event is predicted to occur within the current time interval \([t, t + \Delta t]\), the bacterium moves up to that tumbling time, after which a new direction is sampled. This process repeats until the full time step \(\Delta t\) is completed.

\end{document}